%% file: main.tex
\documentclass[conference,9pt,letterpaper]{IEEEtran}
\IEEEoverridecommandlockouts
\usepackage{cite}
\usepackage{amsmath,amssymb,amsfonts}
\usepackage{algorithmic}
\usepackage{graphicx}
\usepackage{textcomp}
\usepackage{xcolor}
\def\BibTeX{{\rm B\kern-.05em{\sc i\kern-.025em b}\kern-.08em
    T\kern-.1667em\lower.7ex\hbox{E}\kern-.125emX}}
\usepackage{amsmath,amssymb,amsfonts}
\usepackage{algorithmic}
\usepackage{graphicx}
\usepackage{textcomp}
\usepackage{graphicx,import}
\usepackage{xcolor}
\usepackage{float}
\usepackage{xifthen}
\usepackage{tikz}
\usepackage{pgfplots}
\usepackage{mathtools}
\usepackage{siunitx}

\makeatletter
\def\ps@IEEEtitlepagestyle{%
  \def\@oddfoot{\mycopyrightnotice}%
  \def\@oddhead{\hbox{}\@IEEEheaderstyle\leftmark\hfil\thepage}\relax
  \def\@evenhead{\@IEEEheaderstyle\thepage\hfil\leftmark\hbox{}}\relax
  \def\@evenfoot{}%
}
\def\mycopyrightnotice{%
  \begin{minipage}{\textwidth}
  \centering \scriptsize
  Copyright~\copyright~2023 IEEE. Personal use of this material is permitted.  Permission from IEEE must be obtained for all other uses, in any current or future media, including reprinting/republishing this material for advertising or promotional purposes, creating new collective works, for resale or redistribution to servers or lists, or reuse of any copyrighted component of this work in other works. DOI:{10.1109/SSP53291.2023.10208060}
  \end{minipage}
}
\makeatother

\makeatletter
\let\old@ps@headings\ps@headings
\let\old@ps@IEEEtitlepagestyle\ps@IEEEtitlepagestyle
\def\confheader#1{%
  \def\ps@headings{%
    \old@ps@headings%
    \def\@oddhead{\strut\hfill#1\hfill\strut}%
    \def\@evenhead{\strut\hfill#1\hfill\strut}%
  }%
  \def\ps@IEEEtitlepagestyle{%
    \old@ps@IEEEtitlepagestyle%
    \def\@oddhead{\strut\hfill#1\hfill\strut}%
    \def\@evenhead{\strut\hfill#1\hfill\strut}%
  }%
  \ps@headings%
}
\makeatother

\confheader{%
\begin{minipage}{\textwidth}
  \centering \scriptsize
  This document is a preprint of: F. Gast, S. Zeitz, M. Dörpinghaus and G. P. Fettweis, "Comparing Iterative and Least-Squares Based Phase Noise Tracking in Receivers with 1-bit Quantization and Oversampling," 2023 IEEE Statistical Signal Processing Workshop (SSP), Hanoi, Vietnam, 2023, pp. 31-35, doi: 10.1109/SSP53291.2023.10208060.
  \end{minipage}
}

\usepackage[font=small,skip=5pt]{caption}
\usepackage[shrink=30,letterspace=500]{microtype}

\input{defs_Flo.tex}
\pgfplotsset{compat=1.18}

\begin{document}

\title{Comparing Iterative and Least-Squares Based Phase Noise Tracking in Receivers with 1-bit Quantization and Oversampling}
\author{
	\IEEEauthorblockN{Florian Gast, Stephan Zeitz, Meik Dörpinghaus, and Gerhard P. Fettweis }
	\IEEEauthorblockA{\small Vodafone Chair Mobile Communications Systems, Technische Universität Dresden, Germany }
	\IEEEauthorblockA{\small{\{florian.gast, stephan.zeitz, meik.doerpinghaus, gerhard.fettweis\}@tu-dresden.de}}
	}

\maketitle

\let\thefootnote\relax\footnotetext{This work was supported by the German Federal Ministry of Education and Research (BMBF) (6G-life, project-ID 16KISK001K) and the European Union’s
Horizon 2020 research and innovation programme under grant
agreement No. 101015956 Hexa-X. 
Computations were performed at the Center for Information Services and High Performance Computing (ZIH) at TU Dresden.
\vspace{-0.6cm}
}

\begin{abstract}
High data rates require vast bandwidths, that can be found in the sub-THz band, and high sampling frequencies, which are predicted to lead to a problematically high analog-to-digital converter (ADC) power consumption. It was proposed to use 1-bit ADCs to mitigate this problem. Moreover, oscillator phase noise is predicted to be especially high at sub-THz carrier frequencies. For synchronization the phase must be tracked based on 1-bit quantized observations. We study iterative data-aided phase estimation, i.e., the expectation-maximization and the Fisher-scoring algorithm, compared to least-squares (LS) phase estimation. For phase interpolation at the data symbols, we consider the Kalman filter and the Rauch-Tung-Striebel algorithm. Compared to LS estimation, iterative phase noise tracking leads to a significantly lower estimation error variance at high signal-to-noise ratios. However, its benefit for the spectral efficiency using zero-crossing modulation (ZXM) is limited to marginal gains for high faster-than-Nyquist signaling factors, i.e., higher order ZXM modulation.
\end{abstract}

\section{Introduction}
Future applications will require wireless links with enormous data rates, e.g., for virtual reality headsets with high resolution and high frame rate.
These data rates can potentially be provided by using the vast unused bandwidth available in the sub-THz band. A projected challenge when using this bandwidth is the power consumption of the hardware components. 
Especially the analog-to-digital converter (ADC) is predicted to form a bottleneck at high sampling frequencies as its power consumption increases quadratically with the sampling frequency beyond $\SI{300}{MHz}$ \cite{Murmann}. In order to keep the energy consumption at high bandwidths within acceptable limits, a new approach is necessary. Instead of using high resolution ADCs it was proposed to use temporally oversampled 1-bit ADCs instead \cite{ZXM}. Using 1-bit quantization only allows the observation of zero-crossings of the receive signal. The timing of these zero-crossings is used for information bearing in a modulation scheme called zero-crossing modulation (ZXM) \cite{ZXM}. The combination of faster-than Nyquist (FTN) signaling with run-length-limited sequences (RLL) enables to generate such signals, where the information is encoded in the run-lengths between the zero-crossings \cite{landau20181}. 
This corresponds to a paradigm shift by conveying information in the time-domain instead of the amplitude-domain. It is also suspected that using a 1-bit ADC will allow for additional energy savings by reducing linearity requirements of the analog hardware and can potentially make an automatic gain control obsolete.\looseness-1

Oscillators for carrier frequencies around $\SI{100}{GHz}$ are often based on frequency multiplication circuits. This leads to amplified phase noise, which is also predicted by currently used phase noise models \cite{Hajimiri}. This phase noise needs to be tracked for phase noise compensation to enable  coherent detection. When using FTN, the intentional intersymbol interference makes the system more prone to phase noise and, thus, the receiver requires especially accurate phase estimation.

The problem of tracking phase noise in 1-bit quantized systems with temporal oversampling was first considered in \cite{Gast202205}, where the least-squares (LS) estimator from \cite{LS}, originally derived for the estimation of constant phase offsets, was used for phase estimation at pilot symbol time instants. These LS phase estimates were then interpolated by a Kalman filter or the Rauch-Tung-Striebel (RTS) algorithm to estimate phase noise at data symbol time instants. Recently, the Bayesian Cramér-Rao lower bound (BCRB) for this problem was derived in \cite{Zeit2212:Bayesian} and it was shown that our approach from \cite{Gast202205} works remarkably well in the low-to-mid signal-to-noise ratio (SNR) range. However, as the performance of the LS phase estimates at the pilot symbol positions starts to diverge from the BCRB with increasing SNR, the interpolation algorithms fail to reach the BCRB at mid-to-high SNR. For a constant phase offset, it was shown in \cite{Schl202107} that the LS phase estimate can be improved in the high SNR regime by employing iterative algorithms, namely the expectation-maximization (EM) algorithm \cite{Dempster1977} and the Fisher-scoring algorithm \cite[Section 7.7]{Kay1993}, at the price of increased computational complexity.
These findings motivate to also include iterative phase estimation algorithms in the multi-stage phase noise tracking approach from \cite{Gast202205}. 
In this work, the goal is to study the benefit of iterative algorithms for phase noise tracking in receivers with temporally oversampled 1-bit quantization. Therefore, we evaluate the phase estimation error variance and the spectral efficiency gain for ZXM when including iterative phase estimation algorithms compared to the LS-based phase noise tracking approach presented in \cite{Gast202205}.\looseness-1

The rest of this paper is organized as follows. In Section~\ref{sec:SysModel} we introduce the system model including the phase noise model. In Section~\ref{sec:EstimationStructure} the structure of the estimator and the algorithms used are explained in more detail, before we compare their estimation error variance to the BCRB and evaluate their influence on the system performance in Section~\ref{sec:NumericalEvaluation}.
Finally, the work is concluded in Section~\ref{sec:Conclusion}.

\section{System Model}
\label{sec:SysModel}
The considered system model is equivalent to the model used in~\cite{Gast202205}.
Although we consider ZXM specifically, the estimation algorithms are applicable to arbitrary systems using 1-bit quantization at the receiver.

\begin{figure}
\center
\resizebox{0.45\textwidth}{!}{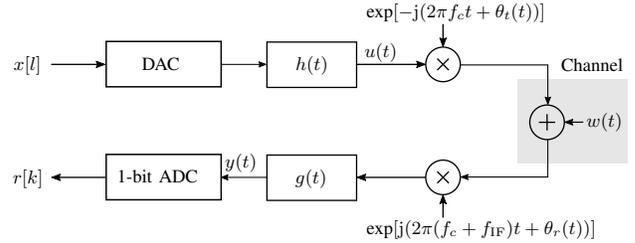} \vspace{0.2cm}
\caption{System model \cite{Gast202205}}
\label{fig:SystemModel}
\end{figure}

\subsection{Transmission Model}
An overview of the system model is depicted in  Fig.~\ref{fig:SystemModel}. 
For transmission we are using QPSK symbols, i.e., 
\begin{equation}
x[l] \in \mathcal{X}= \left\lbrace \frac{1+\mathrm{j}}{\sqrt{2}},\frac{1-\mathrm{j}}{\sqrt{2}},\frac{-1+ \mathrm{j}}{\sqrt{2}},\frac{-1- \mathrm{j}}{\sqrt{2}} \right\rbrace,
\end{equation}
which are stacked in a symbol vector $\mathbf{x}$. The linearly modulated transmit signal is given by
\begin{equation}
u(t)=\sum_{l=0}^{\infty}x[l]h\left(t-\frac{lT}{M_\mathrm{tx}}\right),
\end{equation}
with the transmit filter impulse response $h(t)$. Moreover, $M_{\mathrm{tx}}$ is the FTN signaling factor w.r.t. the Nyquist rate $1/T$.

As described in \cite{ZXM}, RLL sequences are required to control the intentional intersymbol interference when using FTN. The RLL sequences are defined by a $(d,k)$-sequence, consisting of $0$s and $1$s, where two $1$s are separated by at least $d$ and at most $k$ zeros. Together with non-return-to-zero inverted (NRZI) coding this yields RLL sequences consisting of $1$s and $-1$s, where at least $d+1$ consecutive symbols are identical. 
The stream of the complex valued transmit symbols $x[l]$ is formed by using such RLL sequences for the real and the imaginary component separately. The minimum runlength constraint is chosen as $d=M_\mathrm{tx}-$1 \cite{PeterOJCOMS}. Note that the choice of $M_\mathrm{tx}$ defines the grid on which zero-crossings of the transmit signal can be placed.\looseness-1

When upconverting the baseband signal $u(t)$ to the carrier frequency $f_c$, the transmit oscillator adds a time dependent phase $\theta_t(t)$.
Because of the high carrier frequency we assume a line-of-sight channel, which we model as an additive white Gaussian noise (AWGN) channel. In the receiver, the signal is downconverted to the baseband with the receive oscillator. After downconversion the AWGN leads to the white circularly symmetric complex Gaussian noise process $\tilde{w}(t)$ with power spectral density $N_0$.
The receive oscillator again suffers from phase inaccuracies and adds a time dependent phase $\theta_r(t)$.
For the evaluation of the phase noise, only the difference between the oscillator phases is of interest, i.e., $\theta(t)=\theta_r(t)-\theta_t(t)$.
The receive oscillator has a slightly detuned frequency, where a small intermediate frequency $f_\mathrm{IF}$ is added to the carrier frequency $f_c$. 
This intermediate frequency is intentionally used and chosen such that $f_\mathrm{IF}T_s$ is irrational, with $T_s$ being the sampling period, ensuring that the error variance of the phase estimate is independent of the phase itself \cite{Schluter2019}. This leads to a rotating receive constellation and, thus, similar to a random phase dither, averages the phase dependent estimation performance.\looseness-1

The downconverted receive signal is filtered with $g(t)$ yielding
\begin{equation}
    y(t)=\int_{-\infty}^{\infty}\left( u(t)e^{\mathrm{j}(2\pi f_\mathrm{IF}t+\theta(t))}+\tilde{w}(t)\right)g(t-\tau) \mathrm{d}\tau,
\end{equation}
which is then sampled at rate $1/T_s$, leading to $y [k]=y(t=kT_s)$ with the oversampling factor $M_\mathrm{
rx}=T/T_s$. Note that $M_\mathrm{rx}\ge M_\mathrm{tx}$ to be able to resolve the zero-crossing time instants of the transmit signal.\looseness-1

As all considered estimation algorithms are data-aided, they require the noise free signal $s[k]=s(t=kT_s)$ for the pilot blocks, where $s(t)$ is equivalent to $y(t)$ without ${w}(t)$ and $\theta(t)$.\looseness-1

The ADC then quantizes the real and the imaginary component of $y[k]$ separately with a resolution of 1-bit, resulting in 
\begin{equation}
    r[k]=\mathrm{sign}(\Re{y[k]})+\mathrm{j}\,\mathrm{sign}(\,\Im{y[k]}).
\end{equation}

For the derivation of the algorithms we assume a rectangular receive filter, as in \cite{Gast202205,LS, Schl202107}. If the bandwidth of this filter is matched to the sampling rate, i.e., when the one-sided bandwidth $W_g=\frac{1}{2 T_s}$, the filtered and sampled noise $\tilde{w}[k]$ becomes white, which allows for mathematically tractable expressions.
When evaluating the system performance, we choose $h(t)$ and $g(t)$ to be root-raised cosine (RRC) filters of single-sided bandwidth $W_h=W_g=\frac{1+\alpha}{2T}$, where $\alpha$ is the rolloff factor, which leads to colored noise in case of oversampling.
\subsection{Phase Noise Model}
As in \cite{Gast202205} and \cite{Zeit2212:Bayesian} we use the phase noise model by Khanzadi et al. \cite{Khanzadi}, which models the time-dependent phase as the superposition of three individual phase noise components $\theta_0$, $\theta_2$, and $\theta_3$, 
such that the sampled phase noise $\theta[k]$ is modelled as
\begin{equation}
    \theta[k]=\theta_0[k]+\theta_2[k]+\theta_3[k].
\end{equation}
These components are characterized by their power spectral densities (PSD) at a frequency offset $f_m$ from the carrier frequency, given as $K_3/f_m^3$, $K_2/f_m^2$, and $K_0$, respectively, i.e.,  $\theta_0$ models the white phase noise, while  $\theta_2$ and $\theta_3$ model the integrated white and integrated flicker noise, respectively. 
As in \cite{Gast202205} we omit the cubic component $\theta_3$, which is assumed to be suppressed by a phase locked loop. 
The white phase noise is distributed as $\theta_0[k]\sim~\mathcal{N}\left(0,2 K_0 W_g \right)$, and $\theta_2[k]$ forms a Wiener process and can thus be modelled as $\theta_2[k]=\theta_2[k-1]+\zeta_2[k]$, where the increments $\zeta_2[k]\sim \mathcal{N}\left(0,4K_2\pi^2 T_s\right)$ are i.i.d., and $\theta_2[-1]=\theta_\mathrm{init}$, with $\theta_\mathrm{init}$ being an arbitrary initial phase.

\section{Estimator Structure}
\label{sec:EstimationStructure}

\begin{figure}
\center
\resizebox{0.45\textwidth}{!}{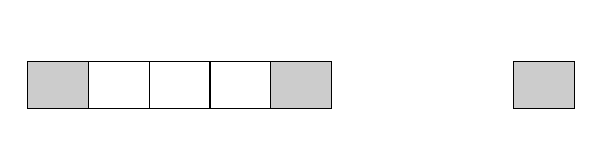}
\caption{Structure of the transmit signal}
\label{fig:TimeSlotModel}
\end{figure}

Similar to \cite{Gast202205} we define a periodic block-based transmit signal with pilot and data blocks as illustrated in Fig.~\ref{fig:TimeSlotModel}.
For the pilot blocks, which consist of $P$ known symbols, we assume $M_{\mathrm{tx}}=1$, since FTN signaling is not needed for the pilot-based phase estimation. Moreover, we consider $M_\mathrm{tx}=M_\mathrm{rx}$ for the $D$ data symbols between two pilot blocks.\looseness-1

As we want to perform interpolation in a block-based manner and, thus, would like to have pilot and data blocks of equal duration, we divide the data transmission between two pilot blocks into blocks of $PM_\mathrm{tx}$ samples, which leads to 
$ 
    \Lambda=\frac{D}{M_\mathrm{tx}P}
$
data blocks of equal length, where we assume $\Lambda$ to be an integer. In the following we use the index $m$ for the blocks.

We further assume the phase noise process to have a small bandwidth compared to the signaling bandwidth, meaning it is valid to assume a constant phase within a pilot block, which justifies the use of estimators derived for constant phase offsets.
The block-based approach is mainly necessary because of the 1-bit quantization, which does not allow sufficiently accurate estimates based on one sample only. Thus, we consider pilot blocks with $P$ pilot symbols.\looseness-1

\begin{figure}
\center
\resizebox{0.47\textwidth}{!}{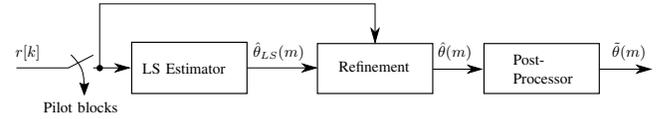}
\caption{Estimator structure}
\label{fig:EstimatorStructure}
\end{figure}

As in \cite{Gast202205}, we use a multi-stage estimator, but insert an additional stage, forming a three-stage estimator, see Fig.~\ref{fig:EstimatorStructure}. While the LS estimator and the iterative algorithms in the refinement step only return an estimate of the phase at the pilot blocks, the interpolation algorithms in the post-processor return a phase estimate for the data blocks as well.\looseness-1

\subsection{LS Estimator}
As in \cite{Schl202107} and \cite{Gast202205}, the LS estimator from \cite{LS} forms the basis of the estimation and is given by
\begin{equation}
    \hat{\theta}_\mathrm{LS}(m)=\mathrm{arg}\left(\sum_{k=mPM_\mathrm{rx}+1}^{(m+1)P M_\mathrm{rx}}s^*[k]r[k]\right),
\end{equation}
where $s^*$ is the complex conjugate of the noise free receive signal $s$, and $m$ needs to be the index of a  pilot block.
\subsection{EM Algorithm}
The EM algorithm, first introduced in \cite{Dempster1977}, is a well-known algorithm to find the maximum likelihood solution in an iterative manner. 
In \cite{Schl202107} it was shown that for the problem of constant phase offset estimation in 1-bit quantized systems it is given as
\begin{equation}
    \hat{\theta}^{l+1}(m)\!=\!\mathrm{arg}\left(\sum_{k=mPM_\mathrm{rx}+1}^{(m+1)P M_\mathrm{rx}} \!\!\!\!\!\!\!\!\!\!\!s^*[k]\!\left(s_{\hat{\theta}}[k]\!+\!\E{\tilde{w}[k]\Big|r[k],\mathbf{x}_m;\hat{\theta}^l(m)} \right)\!\! \right)\!\!, \label{eq:UpdateEM}
\end{equation}
with $s_{\hat{\theta}}[k]=s[k]\exp\!\!\left(j\hat{\theta}^l(m)\right)$ denoting the noise free signal rotated by the last estimate, and $\mathbf{x}_m$ describing the symbols in the $m$-th pilot block. As shown in \cite{mezghani2010multiple}, for white additive noise, i.e., when using a rectangular receive filter with a bandwidth matched to the sampling rate, i.e., $W_g=\frac{1}{2T_s}$, the expectation of the noise in (\ref{eq:UpdateEM}) can be expressed as\looseness-1 
\begin{align}
&\hspace{-1cm}\E{\tilde{w}[k] \Big| r[k], \mathbf{x}_m; \hat{\theta}^l(m)}\nonumber\\
	=& \,\Re{r[k]} \frac{\sigma}{2\sqrt{\pi}} \frac{\exp[-\frac{\Re{s_{\hat{\theta}}[k])}^2}{\sigma^2}]}{\Q{-\Re{r[k]} \frac{\Re{s_{\hat{\theta}}[k])}}{\sigma / \sqrt{2}}}}\nonumber \\
	&+ j\Im{r[k]} \frac{\sigma}{2\sqrt{\pi}} \frac{\exp[-\frac{\Im{s_{\hat{\theta}}[k])}^2}{\sigma^2}]}{\Q{-\Im{r[k]} \frac{\Im{s_{\hat{\theta}}[k])}}{\sigma / \sqrt{2}}}} ,
\end{align}
where $Q(.)$ is the Gaussian Q-function, and $\sigma^2\!=\!N_0/T_s$ is the variance of the filtered white Gaussian noise $\tilde{w}[k]$.
Although this algorithm was derived for white noise and a constant phase offset, it can be applied to the case of phase noise and colored noise. \cite{Schl202107} reported about the slow convergence of the algorithm, which has also been observed when applying the algorithm in a system with phase noise.
\subsection{Fisher-Scoring Algorithm}
The Fisher-Scoring algorithm \cite[Section 7.7]{Kay1993} is a form of the Newton-Raphson method and has a faster convergence than the EM algorithm, although convergence is not guaranteed.
It is defined by\looseness-1
\begin{equation}
    \hat{\theta}^{l+1}(m)=\hat{\theta}^{l}(m)+\mathcal{I}^{-1}(\hat{\theta}^l(m))V(\hat{\theta}^l(m)),\label{eq:UpdateEQScoring}
\end{equation}
where $\mathcal{I}(\cdot)$ denotes the Fisher information and $V(\cdot)$ denotes the score. 
Due to the use of intermediate frequency sampling, the Fisher information is independent of the actual realization of the phase $\hat{\theta}^l(m)$, which allows us to use the lower bound on the inverse of the Fisher information for the case of a constant phase given in  \cite{schluter2020bounds}
\begin{equation}
	\mathcal{I}^{-1}(\hat{\theta}^l(m)) \geq \left( \frac{1}{\pi} \kappa_1\left(\frac{E_s}{N_0 M_\mathrm{rx}}\right) \frac{E_s}{N_0}P\right)^{-1}=	\mathcal{I}^{-1},
\label{eq:FI_inv}
\end{equation}
with
\begin{equation}
	\kappa_1\left(x\right) = c_1 e^{-c_2 x} \left( I_0\left(c_2 x\right) + I_1\left(c_2 x\right) \right),
\end{equation}
where $I_v(\cdot)$ is the modified Bessel function of the first kind, $E_s=\E{|x[l]|^2}\int |h(t)|^2 \mathrm{d}t$,  $c_1=4.0360$, and $c_2=0.3930$.

The score is given as \cite{Schl202107} 
\begin{align}
V(\hat{\theta}^l(m)) 
	&= \!\!\!\!\!\!\!\sum_{k=mPM_\mathrm{rx}+1}^{(m+1)P M_\mathrm{rx}} \!\!\left( \!\!-\Re{r[k]}\! \frac{\exp[-\frac{\Re{s_{\hat{\theta}}[k]}^2}{\sigma^2}] \!\Im{s_{\hat{\theta}}[k]}}{\sigma \sqrt{\pi}\Q{-\!\Re{r[k]} \frac{\Re{s_{\hat{\theta}}[k]}}{\sigma / \sqrt{2}}}}\! \right. \nonumber \\
	&\left. + \Im{r[k]} \frac{\exp[-\frac{\Im{s_{\hat{\theta}}[k]}^2}{\sigma^2}] \Re{s_{\hat{\theta}}[k]}}{\sigma \sqrt{\pi}\Q{-\Im{r[k]} \frac{\Im{s_{\hat{\theta}}[k]}}{\sigma / \sqrt{2}}}} \right).
\end{align}
In the constant phase case, the Fisher-scoring algorithm converged when initialized with the LS estimate, but did not converge in general \cite{Schl202107}. In the presence of phase noise, however, the algorithm does not converge at high SNR, even when initialized rather accurately with the LS estimate. 
To ensure convergence, we modified the algorithm to a damped Newton-Raphson method, by inserting a dampening factor $\epsilon$ in (\ref{eq:UpdateEQScoring}), i.e.,\looseness-1
\begin{equation}
\hat{\theta}^{l+1}(m)=\hat{\theta}^{l}(m)+\epsilon\,\mathcal{I}^{-1}V(\hat{\theta}^l(m)).
\end{equation}
The parameter $\epsilon$ has a strong impact on the convergence speed but ensures convergence in our numerical evaluations.


\subsection{Interpolation}
\begin{figure}
\center
\vspace{.2cm}
\resizebox{0.43\textwidth}{!}{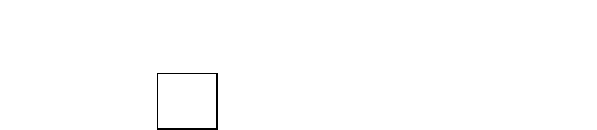}
\caption{Observation model for interpolation \cite{Gast202205}\vspace{-0.2cm}}
\label{fig:ObservationModel}
\end{figure}

In \cite{Gast202205} it was shown that in the presence of phase noise, a block-based tracking with the LS estimator leads to tracking of the sampled time averaged phase noise (STAPN) for each block $m$, given by\looseness-1
\begin{equation}
    \bar{\theta}(m)=\frac{1}{PM_\mathrm{rx}}\sum_{k=mPM_\mathrm{rx}+1}^{(m+1)PM_\mathrm{rx}}\left(\theta_2[k]+\theta_0[k] \right). \label{eq:defSTAPN}
\end{equation}
Similarly, the iterative algorithms track the STAPN as well, although with a lower estimation error variance at high SNR. Hence, for the pilot block phase estimates of all three estimation algorithms (LS, EM, and Fisher-scoring) it holds that
\begin{equation}
    \hat{\theta}(m)=\bar{\theta}(m)+\eta(m),
\end{equation}
with different variances of  $\eta(m)$.
For the interpolation algorithms we model the estimation error $\eta(m)$ to be zero-mean Gaussian with a variance equal to the inverse Fisher-Information from (\ref{eq:FI_inv}), which lower-bounds the estimation error variance of all three estimators.
In \cite{Gast202205} it was also shown that the STAPN can be modeled as a Wiener process similar to its underlying process $\theta_2$,
leading to 
\begin{equation}
    \bar{\theta}(m)=\bar{\theta}(m-1)+\Delta(m), \label{eq:STAPNIncDef}
\end{equation}
where we model the innovation process as Gaussian with $\Delta(m) \sim \mathcal{N}\left(0,\frac{2}{3}PM_\mathrm{rx}\var{\zeta_2[k]} \right)$ \cite{Gast202205}. 
Due to the observations of the phase at pilot blocks only, this leads to an observation model as shown in Fig.~\ref{fig:ObservationModel}.

In \cite{Gast202205} two interpolation algorithms were presented to provide a phase estimate at the data blocks based on the observation model from Fig.~\ref{fig:ObservationModel}: the Kalman-Filter and the Rauch-Tung-Striebel (RTS) algorithm \cite{rauch1965maximum}. The latter one consists of a forward filtering sweep, which is equivalent to the Kalman filter, and a backward smoothing sweep.
It was shown that the RTS algorithm outperforms the Kalman filter at the price of non-causality and, thus, increased latency.\looseness-1
\section{Numerical Evaluation}
\label{sec:NumericalEvaluation}

\subsection{Estimation Error Variance}
\label{sec:EstimationError}
To evaluate the phase estimation performance, we numerically evaluated the phase estimation error variance for a transmit sequence consisting of $K$ pilot blocks and $(K-1)\Lambda$ data blocks. To efficiently use the RTS algorithm we assume each sequence to start and end with a pilot block. Since we assume $M_\mathrm{rx}=M_\mathrm{tx}$ for the data blocks, this leads to $N=KPM_\mathrm{rx}+(K-1)D$ samples.\looseness-1

The estimation error variance of the algorithms is compared by evaluating the sample-based mean squared estimation error averaged over the entire observation interval, i.e.,
\begin{equation}
    \bar{\xi}=\frac{1}{N}\sum_{k=1}^{N}\left(\theta[k]-\tilde{\theta}(m) \right)^2,
\end{equation} 
where $\tilde{\theta}(m)$ is the output of the multi-stage estimator, i.e., of the interpolator, as shown  in Fig.~\ref{fig:EstimatorStructure} for the block $m$ containing the sample $k$.\looseness-1

\begin{figure}
\center
\resizebox{0.45\textwidth}{!}{\input{figures/WhiteNoise+BCRB}}
\caption{Estimation error variance and BCRB for $M_\mathrm{rx}=M_\mathrm{tx}=1,D=180, P=60, K_2=800, K_0=-130\,\mathrm{dB}$, $\epsilon=0.05$ and a rectangular receive filter with $W_g=1/(2T_s)$ (white noise). EM and Fisher-scoring algorithm both used 20 iterations.}
\label{fig:MSE_WhiteNoise+BCRB}
\end{figure}
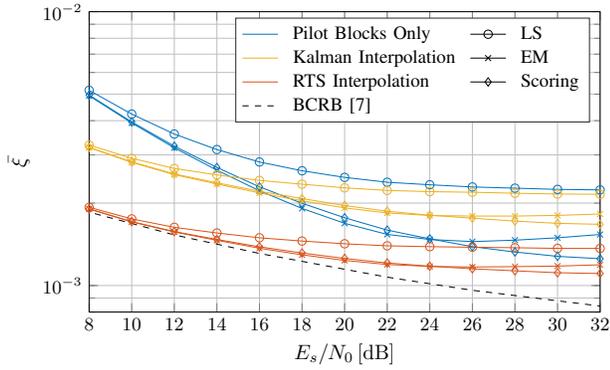
Fig.~\ref{fig:MSE_WhiteNoise+BCRB} shows the estimation error variance $\bar{\xi}$ for the case $M_\mathrm{tx}=M_\mathrm{rx}=1$, since this setting allows for comparison with the BCRB from \cite{Zeit2212:Bayesian}. For a comparison of the performance of the LS, the Fisher-scoring, and the EM algorithm, without the effect of interpolation, we evaluate their estimation error variance also at pilot blocks only.
It can be seen that the iterative algorithms improve the LS estimate, which also leads to a better tracking performance of both the Kalman filter and the RTS algorithm. The interpolation algorithms using the iterative phase estimates still reach an error floor. However, this error floor is significantly lower than when using the LS estimate without refinement. Interestingly, the estimation performance of the EM estimator starts to deteriorate again at very high $E_s/N_0$, which is due to the fixed number of iterations as the EM algorithm's convergence speed is dependent on the noise variance, as elaborated in \cite{Schl202107}. Comparing the RTS interpolated estimates of the iterative algorithms and of the LS estimator with the BCRB shows that the estimates of the iterative algorithms lead to a significant performance gain and stay closer to the BCRB.

\subsection{System Performance}
To evaluate if the additional step of utilizing iterative algorithms is worth the computational effort and the increase in latency, the influence of the phase estimation on the system performance needs to be evaluated.
For this we employ a system nearly identical to the one described in \cite{PeterOJCOMS} using finite-state machine based RLL codes with a minimal run-length of $d=M_\mathrm{tx}-1$.
The presented algorithms are used to track the phase, which is corrected in the analog domain. Then a lower bound on the achievable spectral efficiency (SE) is evaluated by empirically estimating the mutual information between the RLL encoder input bits and the log-likelihood ratios at the RLL decoder output as in \cite[Section V]{PeterOJCOMS}.

\begin{figure}
\center
\resizebox{0.45\textwidth}{!}{\input{figures/SE_bundle_curves}}
\caption{Spectral efficiency (SE) lower bound using LS or Fisher-scoring based pilot block phase estimates succeeded by Kalman or RTS interpolation for different FTN signaling factors $M_\mathrm{tx}$. For all settings: $K_2=1200, K_0=-130\,\mathrm{dB}$, $M_\mathrm{rx}=M_\mathrm{tx}$ for data blocks, RRC transmit and receive filter with $W_h=W_g=\frac{1+\alpha}{2T}$, rolloff factor $\alpha=0.6$, $\epsilon=0.4$, 20 scoring iterations, $P=30$ symbols, $D/M_\mathrm{tx}=600$.}
\label{fig:SE_bundles}
\end{figure}
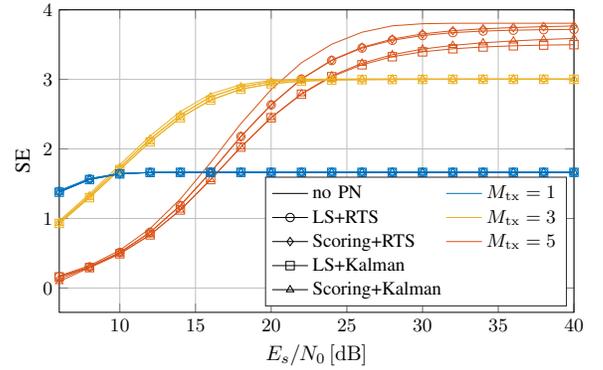

Fig. \ref{fig:SE_bundles} shows the spectral efficiency lower bound for different phase estimation and interpolation algorithms, and FTN signaling factors. The presented results take the decrease of the spectral efficiency due to the pilot symbols into account. Since the performance of the EM algorithm and the Fisher-scoring algorithm is almost identical, we only show the latter one. It becomes clear that the iterative algorithms only significantly improve the SE at high $E_s/N_0$ for high $M_\mathrm{tx}$.
For $M_\mathrm{tx}=1$ and $M_\mathrm{tx}=3$ there is no significant difference between the different tracking approaches and the scenario without phase noise at all.
For $M_\mathrm{tx}=5$ it can be seen, that phase noise causes a notable spectral efficiency degradation. Overall, RTS interpolation leads to a significantly higher SE compared to Kalman interpolation. However, the displayed SE results of the RTS algorithm are based on a non-causal phase correction in the analog domain and, thus, correspond to an upper bound. For the Kalman interpolation the difference in SE of the LS-based and iteratively refined approach is at maximum about $2.5\,\%$ and, thus, the computational effort is likely not necessary.

\section{Conclusion}
\label{sec:Conclusion}
In this paper we showed that using the computationally inexpensive, and thus low-latency, LS phase estimator is sufficient in most scenarios and that using the iterative algorithms from \cite{Schl202107} as an additional stage in the multi-stage phase noise estimator from \cite{Gast202205} is likely not needed in a practical system. To show this, we compared the two-stage estimator from \cite{Gast202205} with a three-stage estimator, utilizing iterative phase estimation algorithms using the LS estimator as initialization. While the estimation error variance of the three-stage estimator is significantly lower and stays closer to the BCRB than the one of the previously available two-stage estimator, the system performance evaluations for ZXM showed that the improved phase estimates do not lead to significant performance gains for the considered  FTN signaling factors. Although we performed our numerical evaluations based on ZXM, the algorithms are also applicable for systems with receivers using 1-bit quantization and different waveforms, where the SE gains of the presented three-stage estimator might be higher.

\bibliographystyle{IEEEtran}
\bibliography{Bibliography}
\end{document}

%% file: defs_Flo.tex
\renewcommand{\Re}[1]{\mathrm{Re}\left\{#1\right\}}
\renewcommand{\Im}[1]{\mathrm{Im}\left\{#1\right\}}
\newcommand{\Q}[1]{Q\left(#1\right)}

\newcommand{\E}[2][]
{
	\ifthenelse{\isempty{#1}}
		{\mathbb{E}\left[#2\right]}
		{\mathbb{E}_{#1}\left[#2\right]}
}

\newcommand{\var}[2][]
{
	\ifthenelse{\isempty{#1}}
		{\text{Var}\left[#2\right]}
		{\text{Var}_{#1}\left[#2\right]}
}
\newcommand{\rpm}{\sbox0{$1$}\sbox2{$\scriptstyle\pm$}
  \raise\dimexpr(\ht0-\ht2)/2\relax\box2 }
  
\renewcommand{\ln}[1][]
{
	\ifthenelse{\isempty{#1}}
		{\mathrm{ln}\,}
		{\mathrm{ln}\left(#1\right)}
}

\renewcommand{\exp}[1][]
{
	\ifthenelse{\isempty{#1}}
		{\mathrm{exp}\,}
		{\mathrm{exp}\left\{#1\right\}}
}

%% file: SystemModel2.pdf_tex
\begingroup%
  \makeatletter%
  \providecommand\color[2][]{%
    \errmessage{(Inkscape) Color is used for the text in Inkscape, but the package 'color.sty' is not loaded}%
    \renewcommand\color[2][]{}%
  }%
  \providecommand\transparent[1]{%
    \errmessage{(Inkscape) Transparency is used (non-zero) for the text in Inkscape, but the package 'transparent.sty' is not loaded}%
    \renewcommand\transparent[1]{}%
  }%
  \providecommand\rotatebox[2]{#2}%
  \newcommand*\fsize{\dimexpr\f@size pt\relax}%
  \newcommand*\lineheight[1]{\fontsize{\fsize}{#1\fsize}\selectfont}%
  \ifx\svgwidth\undefined%
    \setlength{\unitlength}{296.42538133bp}%
    \ifx\svgscale\undefined%
      \relax%
    \else%
      \setlength{\unitlength}{\unitlength * \real{\svgscale}}%
    \fi%
  \else%
    \setlength{\unitlength}{\svgwidth}%
  \fi%
  \global\let\svgwidth\undefined%
  \global\let\svgscale\undefined%
  \makeatother%
  \begin{picture}(1,0.36727758)%
    \lineheight{1}%
    \setlength\tabcolsep{0pt}%
    \put(0,0){\includegraphics[width=\unitlength,page=1]{SystemModel2.pdf}}%
    \put(0.20877623,0.26839874){\color[rgb]{0,0,0}\makebox(0,0)[lt]{\lineheight{1.25}\smash{\begin{tabular}[t]{l}DAC\end{tabular}}}}%
    \put(0.45888302,0.26792466){\color[rgb]{0,0,0}\makebox(0,0)[lt]{\lineheight{1.25}\smash{\begin{tabular}[t]{l}$h(t)$\end{tabular}}}}%
    \put(0.57469421,0.35478076){\color[rgb]{0,0,0}\makebox(0,0)[lt]{\lineheight{1.25}\smash{\begin{tabular}[t]{l}$\text{exp}[-\mathrm{j}(2\pi f_ct+\theta_t(t))]$\end{tabular}}}}%
    \put(0.57469421,0.00162088){\color[rgb]{0,0,0}\makebox(0,0)[lt]{\lineheight{1.25}\smash{\begin{tabular}[t]{l}$\text{exp}[\mathrm{j}(2 \pi (f_c+f_\mathrm{IF})t+\theta_r(t))]$\end{tabular}}}}%
    \put(0.93290699,0.1761575){\color[rgb]{0,0,0}\makebox(0,0)[lt]{\lineheight{1.25}\smash{\begin{tabular}[t]{l}$w(t)$\end{tabular}}}}%
    \put(0.45850415,0.08430516){\color[rgb]{0,0,0}\makebox(0,0)[lt]{\lineheight{1.25}\smash{\begin{tabular}[t]{l}$g(t)$\end{tabular}}}}%
    \put(0.16838477,0.08477924){\color[rgb]{0,0,0}\makebox(0,0)[lt]{\lineheight{1.25}\smash{\begin{tabular}[t]{l}1-bit ADC\end{tabular}}}}%
    \put(-0.00021128,0.26934705){\color[rgb]{0,0,0}\makebox(0,0)[lt]{\lineheight{1.25}\smash{\begin{tabular}[t]{l}$x[l]$\end{tabular}}}}%
    \put(-0.00155232,0.08448763){\color[rgb]{0,0,0}\makebox(0,0)[lt]{\lineheight{1.25}\smash{\begin{tabular}[t]{l}$r[k]$\end{tabular}}}}%
    \put(0.89175695,0.26627825){\color[rgb]{0,0,0}\makebox(0,0)[lt]{\lineheight{1.25}\smash{\begin{tabular}[t]{l}Channel\end{tabular}}}}%
    \put(0.57129495,0.28991186){\color[rgb]{0,0,0}\makebox(0,0)[lt]{\lineheight{1.25}\smash{\begin{tabular}[t]{l}$u(t)$\end{tabular}}}}%
    \put(0.34403349,0.11009066){\color[rgb]{0,0,0}\makebox(0,0)[lt]{\lineheight{1.25}\smash{\begin{tabular}[t]{l}$y(t)$\end{tabular}}}}%
  \end{picture}%
\endgroup%

%% file: TimeSlotModel.pdf_tex
\begingroup%
  \makeatletter%
  \providecommand\color[2][]{%
    \errmessage{(Inkscape) Color is used for the text in Inkscape, but the package 'color.sty' is not loaded}%
    \renewcommand\color[2][]{}%
  }%
  \providecommand\transparent[1]{%
    \errmessage{(Inkscape) Transparency is used (non-zero) for the text in Inkscape, but the package 'transparent.sty' is not loaded}%
    \renewcommand\transparent[1]{}%
  }%
  \providecommand\rotatebox[2]{#2}%
  \newcommand*\fsize{\dimexpr\f@size pt\relax}%
  \newcommand*\lineheight[1]{\fontsize{\fsize}{#1\fsize}\selectfont}%
  \ifx\svgwidth\undefined%
    \setlength{\unitlength}{288.06616232bp}%
    \ifx\svgscale\undefined%
      \relax%
    \else%
      \setlength{\unitlength}{\unitlength * \real{\svgscale}}%
    \fi%
  \else%
    \setlength{\unitlength}{\svgwidth}%
  \fi%
  \global\let\svgwidth\undefined%
  \global\let\svgscale\undefined%
  \makeatother%
  \begin{picture}(1,0.23521295)%
    \lineheight{1}%
    \setlength\tabcolsep{0pt}%
    \put(0,0){\includegraphics[width=\unitlength,page=1]{TimeSlotModel.pdf}}%
    \put(-0.00297899,0.1851398){\color[rgb]{0,0,0}\makebox(0,0)[lt]{\lineheight{1.25}\smash{\begin{tabular}[t]{l}Pilot\end{tabular}}}}%
    \put(0,0){\includegraphics[width=\unitlength,page=2]{TimeSlotModel.pdf}}%
    \put(0.4642839,0.21043159){\color[rgb]{0,0,0}\makebox(0,0)[lt]{\lineheight{1.25}\smash{\begin{tabular}[t]{l}Data\end{tabular}}}}%
    \put(0,0){\includegraphics[width=\unitlength,page=3]{TimeSlotModel.pdf}}%
    \put(0.2804837,0.00201709){\color[rgb]{0,0,0}\makebox(0,0)[lt]{\lineheight{1.25}\smash{\begin{tabular}[t]{l}$D$\end{tabular}}}}%
    \put(0.07906177,0.00269118){\color[rgb]{0,0,0}\makebox(0,0)[lt]{\lineheight{1.25}\smash{\begin{tabular}[t]{l}$P$\end{tabular}}}}%
    \put(0.00545003,0.09221165){\makebox(0,0)[lt]{\lineheight{1.25}\smash{\begin{tabular}[t]{l}...\end{tabular}}}}%
    \put(0.98677385,0.09122543){\makebox(0,0)[lt]{\lineheight{1.25}\smash{\begin{tabular}[t]{l}...\end{tabular}}}}%
    \put(0,0){\includegraphics[width=\unitlength,page=4]{TimeSlotModel.pdf}}%
  \end{picture}%
\endgroup%

%% file: BlockEstimation.pdf_tex
\begingroup%
  \makeatletter%
  \providecommand\color[2][]{%
    \errmessage{(Inkscape) Color is used for the text in Inkscape, but the package 'color.sty' is not loaded}%
    \renewcommand\color[2][]{}%
  }%
  \providecommand\transparent[1]{%
    \errmessage{(Inkscape) Transparency is used (non-zero) for the text in Inkscape, but the package 'transparent.sty' is not loaded}%
    \renewcommand\transparent[1]{}%
  }%
  \providecommand\rotatebox[2]{#2}%
  \newcommand*\fsize{\dimexpr\f@size pt\relax}%
  \newcommand*\lineheight[1]{\fontsize{\fsize}{#1\fsize}\selectfont}%
  \ifx\svgwidth\undefined%
    \setlength{\unitlength}{375.37000436bp}%
    \ifx\svgscale\undefined%
      \relax%
    \else%
      \setlength{\unitlength}{\unitlength * \real{\svgscale}}%
    \fi%
  \else%
    \setlength{\unitlength}{\svgwidth}%
  \fi%
  \global\let\svgwidth\undefined%
  \global\let\svgscale\undefined%
  \makeatother%
  \begin{picture}(1,0.17065605)%
    \lineheight{1}%
    \setlength\tabcolsep{0pt}%
    \put(0,0){\includegraphics[width=\unitlength,page=1]{BlockEstimation.pdf}}%
    \put(0.19679764,0.06168872){\color[rgb]{0,0,0}\makebox(0,0)[lt]{\lineheight{1.25}\smash{\begin{tabular}[t]{l}LS Estimator\end{tabular}}}}%
    \put(0.77472357,0.07555518){\color[rgb]{0,0,0}\makebox(0,0)[lt]{\lineheight{1.25}\smash{\begin{tabular}[t]{l}Post-\\Processor\end{tabular}}}}%
    \put(-0.00161875,0.08769575){\color[rgb]{0,0,0}\makebox(0,0)[lt]{\lineheight{1.25}\smash{\begin{tabular}[t]{l}$r[k]$\end{tabular}}}}%
    \put(0.36915336,0.08756151){\color[rgb]{0,0,0}\makebox(0,0)[lt]{\lineheight{1.25}\smash{\begin{tabular}[t]{l}$\hat{\theta}_{LS}(m)$\end{tabular}}}}%
    \put(0.92913265,0.08722493){\color[rgb]{0,0,0}\makebox(0,0)[lt]{\lineheight{1.25}\smash{\begin{tabular}[t]{l}$\tilde{\theta}(m)$\end{tabular}}}}%
    \put(0,0){\includegraphics[width=\unitlength,page=2]{BlockEstimation.pdf}}%
    \put(0.50299165,0.063567){\color[rgb]{0,0,0}\makebox(0,0)[lt]{\lineheight{1.25}\smash{\begin{tabular}[t]{l}Refinement\end{tabular}}}}%
    \put(0.65740079,0.08722493){\color[rgb]{0,0,0}\makebox(0,0)[lt]{\lineheight{1.25}\smash{\begin{tabular}[t]{l}$\hat{\theta}(m)$\end{tabular}}}}%
    \put(0,0){\includegraphics[width=\unitlength,page=3]{BlockEstimation.pdf}}%
    \put(0.04240474,0.00028642){\color[rgb]{0,0,0}\makebox(0,0)[lt]{\lineheight{1.25}\smash{\begin{tabular}[t]{l}Pilot blocks\end{tabular}}}}%
    \put(0,0){\includegraphics[width=\unitlength,page=4]{BlockEstimation.pdf}}%
  \end{picture}%
\endgroup%

%% file: ObservationModelPayload.pdf_tex
\begingroup%
  \makeatletter%
  \providecommand\color[2][]{%
    \errmessage{(Inkscape) Color is used for the text in Inkscape, but the package 'color.sty' is not loaded}%
    \renewcommand\color[2][]{}%
  }%
  \providecommand\transparent[1]{%
    \errmessage{(Inkscape) Transparency is used (non-zero) for the text in Inkscape, but the package 'transparent.sty' is not loaded}%
    \renewcommand\transparent[1]{}%
  }%
  \providecommand\rotatebox[2]{#2}%
  \newcommand*\fsize{\dimexpr\f@size pt\relax}%
  \newcommand*\lineheight[1]{\fontsize{\fsize}{#1\fsize}\selectfont}%
  \ifx\svgwidth\undefined%
    \setlength{\unitlength}{286.40764116bp}%
    \ifx\svgscale\undefined%
      \relax%
    \else%
      \setlength{\unitlength}{\unitlength * \real{\svgscale}}%
    \fi%
  \else%
    \setlength{\unitlength}{\svgwidth}%
  \fi%
  \global\let\svgwidth\undefined%
  \global\let\svgscale\undefined%
  \makeatother%
  \begin{picture}(1,0.21773093)%
    \lineheight{1}%
    \setlength\tabcolsep{0pt}%
    \put(0,0){\includegraphics[width=\unitlength,page=1]{ObservationModelPayload.pdf}}%
    \put(0.00119698,0.20328477){\color[rgb]{0,0,0}\makebox(0,0)[lt]{\lineheight{1.25}\smash{\begin{tabular}[t]{l}$\Delta(m)$\end{tabular}}}}%
    \put(0,0){\includegraphics[width=\unitlength,page=2]{ObservationModelPayload.pdf}}%
    \put(0.27987745,0.03434376){\color[rgb]{0,0,0}\makebox(0,0)[lt]{\lineheight{1.25}\smash{\begin{tabular}[t]{l}$z^{-1}$\end{tabular}}}}%
    \put(0.76832172,0.04469807){\color[rgb]{0,0,0}\makebox(0,0)[lt]{\lineheight{1.25}\smash{\begin{tabular}[t]{l}$\eta(m)$\end{tabular}}}}%
    \put(0.4515023,0.20202843){\color[rgb]{0,0,0}\makebox(0,0)[lt]{\lineheight{1.25}\smash{\begin{tabular}[t]{l}$\bar{\theta}(m)$\end{tabular}}}}%
    \put(0.89299003,0.2022302){\color[rgb]{0,0,0}\makebox(0,0)[lt]{\lineheight{1.25}\smash{\begin{tabular}[t]{l}$\hat{\theta}(m)$\end{tabular}}}}%
    \put(0,0){\includegraphics[width=\unitlength,page=3]{ObservationModelPayload.pdf}}%
    \put(0.52638946,0.11611788){\color[rgb]{0,0,0}\makebox(0,0)[lt]{\lineheight{1.25}\smash{\begin{tabular}[t]{l}Pilot blocks\end{tabular}}}}%
  \end{picture}%
\endgroup%

%% file: figures/WhiteNoise+BCRB.tex
\definecolor{mycolor1}{rgb}{0.00000,0.44706,0.74118}%
\definecolor{mycolor2}{rgb}{0.92941,0.69412,0.12549}%
\definecolor{mycolor3}{rgb}{0.85098,0.32549,0.09804}%
\begin{tikzpicture}

\begin{axis}[%
width=3.4in,
height=2in,
at={(0.758in,0.516in)},
scale only axis,
xmin=8,
xmax=32,
xlabel style={font=\color{white!15!black}},
xlabel={$E_s/N_0\,[\mathrm{dB}]$},
ymode=log,
ymin=0.0008,
ymax=0.01,
yminorticks=true,
label style={font=\large},
ylabel={$\bar{\xi}$},
axis background/.style={fill=white},
legend style={legend cell align=left, align=left, draw=white!15!black, draw opacity=1, text opacity=1, column sep=5pt},
legend columns=2,
yminorgrids=true,
xmajorgrids=true
]
\addplot[color=mycolor1,draw=none]
table[row sep=crcr]{%
1 1\\
};\addlegendentry{Pilot Blocks Only}

\addplot [color=black, mark=o, mark options={solid, black} ,draw=none]
table[row sep=crcr]{%
1 1\\
};\addlegendentry{LS}

\addplot[color=mycolor2,draw=none]
table[row sep=crcr]{%
1 1\\
};\addlegendentry{Kalman Interpolation}

\addplot [color=black, mark=x, mark options={solid, black} ,draw=none]
table[row sep=crcr]{%
1 1\\
};\addlegendentry{EM}

\addplot[color=mycolor3,draw=none]
table[row sep=crcr]{%
1 1\\
};\addlegendentry{RTS Interpolation}

\addplot [color=black, mark=diamond, mark options={solid, black} ,draw=none]
table[row sep=crcr]{%
1 1\\
};\addlegendentry{Scoring}

\addplot[color=black,draw=none, dashed]
table[row sep=crcr]{%
1 1\\
};\addlegendentry{BCRB \cite{Zeit2212:Bayesian}}

\addplot [color=mycolor1, mark=o, mark options={solid, mycolor1}]
  table[row sep=crcr]{%
6	0.00644336890898932\\
8	0.00516445183992765\\
10	0.00422812551123255\\
12	0.0035756572038213\\
14	0.00313748780707922\\
16	0.00282574703661955\\
18	0.00262531572062858\\
20	0.00248296869450464\\
22	0.00238378109829297\\
24	0.00233319925290508\\
26	0.00229098647174681\\
28	0.00226765994535559\\
30	0.00224438847181796\\
32	0.00223475382270671\\
34	0.00222576739713225\\
36	0.00221499796400137\\
38	0.002213608507324\\
40	0.0022103699633132\\
};

\addplot [color=mycolor1, mark=x, mark options={solid, mycolor1}]
  table[row sep=crcr]{%
6	0.00626820092383825\\
8	0.0049346103355927\\
10	0.00392575986762381\\
12	0.00318359484234398\\
14	0.00263867151673745\\
16	0.00221824817364527\\
18	0.00191250143987112\\
20	0.00168734513142884\\
22	0.00153743427596204\\
24	0.00147305563847011\\
26	0.00144715165772664\\
28	0.00146457353134727\\
30	0.00149429801944751\\
32	0.00153617246025483\\
34	0.00157057233901057\\
36	0.00159251080774787\\
38	0.00161895914276801\\
40	0.0016342299366971\\
};

\addplot [color=mycolor1, mark=diamond, mark options={solid, mycolor1}]
  table[row sep=crcr]{%
6	0.00628660590294019\\
8	0.0049585496351551\\
10	0.00395864593981668\\
12	0.00322832487567758\\
14	0.00269949715143648\\
16	0.00229378969386827\\
18	0.00199750585518812\\
20	0.00176841844250474\\
22	0.00159395340990366\\
24	0.00148122975487436\\
26	0.00138679604306911\\
28	0.00132578783035358\\
30	0.00127835835889902\\
32	0.00125275142128259\\
34	0.00122863761330313\\
36	0.00120684256972833\\
38	0.00120126882921943\\
40	0.00119432634255563\\
};

\addplot [color=mycolor2, mark=o, mark options={solid, mycolor2}]
  table[row sep=crcr]{%
6	0.00362020852412216\\
8	0.00325801251994071\\
10	0.00291627283621355\\
12	0.00267703449479856\\
14	0.00253541754695455\\
16	0.00242053264761219\\
18	0.00234462932577503\\
20	0.00227429917181469\\
22	0.0022248063827449\\
24	0.00220260244130169\\
26	0.00219098779025051\\
28	0.00217285420520273\\
30	0.00215909746924723\\
32	0.002154149000564\\
34	0.00214858712871729\\
36	0.00214384869453494\\
38	0.00214479072515126\\
40	0.00214156841798985\\
};

\addplot [color=mycolor2, mark=x, mark options={solid, mycolor2}]
  table[row sep=crcr]{%
6	0.00357373359497129\\
8	0.00319233934688447\\
10	0.00281688299893022\\
12	0.00254094655916151\\
14	0.00234889419894071\\
16	0.00217870687451998\\
18	0.00203930976318108\\
20	0.00192124008333567\\
22	0.00184066651984504\\
24	0.00180200858529537\\
26	0.00179362798466485\\
28	0.00179372712254709\\
30	0.00180363546796013\\
32	0.00182480174939789\\
34	0.0018411091880943\\
36	0.00185141840269765\\
38	0.00186510586212649\\
40	0.0018716705644783\\
};

\addplot [color=mycolor2, mark=diamond, mark options={solid, mycolor2}]
  table[row sep=crcr]{%
6	0.00357850658912529\\
8	0.00319868620703603\\
10	0.0028286119893698\\
12	0.002556558720576\\
14	0.00237200043852635\\
16	0.00220956274942265\\
18	0.002078145085062\\
20	0.00195999192405773\\
22	0.00186850882254988\\
24	0.00180631815003328\\
26	0.00176179417197951\\
28	0.00171963797705899\\
30	0.00168730151566365\\
32	0.00167315328758236\\
34	0.00165819367582969\\
36	0.00164415010201605\\
38	0.00163965103604603\\
40	0.00163674050486461\\
};

\addplot [color=mycolor3, mark=o, mark options={solid, mycolor3}]
  table[row sep=crcr]{%
6	0.00212292660966414\\
8	0.00193082602155516\\
10	0.00175167380936344\\
12	0.00163160097381027\\
14	0.00155571526907164\\
16	0.00149409585704667\\
18	0.00145369213619599\\
20	0.00142084448637711\\
22	0.00139664246729344\\
24	0.0013867594297784\\
26	0.00137984193274504\\
28	0.00137425275546301\\
30	0.0013666150320668\\
32	0.00136624342597962\\
34	0.00136520230412778\\
36	0.00136295185110127\\
38	0.00136089736813737\\
40	0.00135938588416116\\
};

\addplot [color=mycolor3, mark=x, mark options={solid, mycolor3}]
  table[row sep=crcr]{%
6	0.0021071342564869\\
8	0.00190389083297961\\
10	0.0017081194269138\\
12	0.00156896917356074\\
14	0.00146283530852277\\
16	0.00136888447001436\\
18	0.00129481175036722\\
20	0.00123523104714517\\
22	0.00119385013483214\\
24	0.00117426659039635\\
26	0.00116869020891901\\
28	0.00117245527442125\\
30	0.0011776055753556\\
32	0.00119064250311098\\
34	0.00120107193409331\\
36	0.00120729688295199\\
38	0.00121341276400477\\
40	0.00121702927640418\\
};

\addplot [color=mycolor3, mark=diamond, mark options={solid, mycolor3}]
  table[row sep=crcr]{%
6	0.00210692381744773\\
8	0.00190497709446087\\
10	0.0017114007690316\\
12	0.00157422910624163\\
14	0.00147356460368218\\
16	0.00138498640524101\\
18	0.00131520651834321\\
20	0.00125606346943446\\
22	0.00120884097575702\\
24	0.00117663554892423\\
26	0.0011512364156725\\
28	0.00113184238260261\\
30	0.00111365259309332\\
32	0.00110644106991093\\
34	0.00109954237659196\\
36	0.00109282038316309\\
38	0.00109076805125544\\
40	0.00108899131035037\\
};

\addplot [color=black, dashed]
  table[row sep=crcr]{%
6	0.00203945582765004\\
12	0.00153100151222062\\
16	0.00130954360324524\\
22	0.00107306536948525\\
26	0.000960954116656178\\
32	0.000841599065982139\\
36	0.000785921020962587\\
};

\end{axis}

\end{tikzpicture}%

%% file: figures/SE_bundle_curves.tex
\definecolor{mycolor1}{rgb}{0.00000,0.44706,0.74118}%
\definecolor{mycolor2}{rgb}{0.92941,0.69412,0.12549}%
\definecolor{mycolor3}{rgb}{0.85098,0.32549,0.09804}%

\usetikzlibrary{spy}
\begin{tikzpicture}[spy using outlines=
	{circle, magnification=3, connect spies}]

\begin{axis}[%
width=3.4in,
height=2in,
at={(3.467in,1.521in)},
scale only axis,
xmin=6,
xmax=40,
xlabel={$E_s/N_0\,[\mathrm{dB}]$},
ymin=-0.35,
ymax=4 ,
ylabel style={font=\color{white!15!black}},
ylabel={SE},
axis background/.style={fill=white},
legend columns=2,
label style={font=\large},
legend style={at={(0.4,0.02)}, anchor=south west, legend cell align=left, align=left, draw=white!15!black},
ymajorgrids=true,
xmajorgrids=true
]

\addplot [color=black, ,draw=none]
table[row sep=crcr]{%
1 1\\
};\addlegendentry{no PN}

\addplot[color=mycolor1,draw=none]
table[row sep=crcr]{%
1 1\\
};\addlegendentry{$M_{\mathrm{tx}}=1$}

\addplot [color=black, mark=o, mark options={solid, black} ,draw=none]
table[row sep=crcr]{%
1 1\\
};\addlegendentry{LS+RTS}

\addplot[color=mycolor2,draw=none]
table[row sep=crcr]{%
1 1\\
};\addlegendentry{$M_{\mathrm{tx}}=3$}

\addplot [color=black, mark=diamond, mark options={solid, black} ,draw=none]
table[row sep=crcr]{%
1 1\\
};\addlegendentry{Scoring+RTS}

\addplot[color=mycolor3,draw=none]
table[row sep=crcr]{%
1 1\\
};\addlegendentry{$M_{\mathrm{tx}}=5$}
\addplot [color=black, mark=square, mark options={solid, black} ,draw=none]
table[row sep=crcr]{%
1 1\\
};\addlegendentry{LS+Kalman}

\addlegendimage{empty legend}\addlegendentry{} 

\addplot [color=black, mark=triangle, mark options={solid, black} ,draw=none]
table[row sep=crcr]{%
1 1\\
};\addlegendentry{Scoring+Kalman}

\def\fac{0.952380}

\addplot [color=mycolor3]
  table[row sep=crcr , y expr=\thisrowno{1}*\fac]{%
6	0.182744499969597\\
8	0.337741034979484\\
10	0.570275669664829\\
12	0.897833970164158\\
14	1.34291167357375\\
16	1.89435667261282\\
18	2.482787131778\\
20	2.99661558915722\\
22	3.40146944034495\\
24	3.68089719016379\\
26	3.8619389242841\\
28	3.95821347150482\\
30	3.99035861644625\\
32	3.99718990912835\\
34	3.99763792502212\\
36	3.99763792502211\\
38	3.99763792502212\\
40	3.99763792502211\\
};

\addplot [color=mycolor3, mark=square, mark options={solid, mycolor3}]
  table[row sep=crcr, y expr=\thisrowno{1}*\fac]{%
6	0.167976218643968\\
8	0.307822361586414\\
10	0.516783237820378\\
12	0.80247141978529\\
14	1.17823035491337\\
16	1.63706198822456\\
18	2.12762601018814\\
20	2.57081915756927\\
22	2.92639926887008\\
24	3.19010290528652\\
26	3.37073633458273\\
28	3.49199078357404\\
30	3.56544840687484\\
32	3.61193907855432\\
34	3.63897957258747\\
36	3.65664526126233\\
38	3.66752943191188\\
40	3.67355234544919\\
};

\addplot [color=mycolor3, mark=o, mark options={solid, mycolor3}]
  table[row sep=crcr , y expr=\thisrowno{1}*\fac]{%
6	0.173133377645648\\
8	0.318511571694404\\
10	0.536836962317233\\
12	0.841278348625763\\
14	1.24659739795566\\
16	1.74763810463363\\
18	2.28742431971446\\
20	2.7656234568479\\
22	3.15292943131287\\
24	3.4352115769496\\
26	3.62315380875054\\
28	3.73990296406524\\
30	3.81240418397335\\
32	3.85540129192167\\
34	3.87951406708635\\
36	3.8924287540571\\
38	3.90024656147434\\
40	3.90459613424116\\
};

\addplot [color=mycolor3, mark=triangle, mark options={solid, mycolor3}]
  table[row sep=crcr , y expr=\thisrowno{1}*\fac]{%
6	0.098997229023723\\
8	0.307582798613617\\
10	0.516662643378521\\
12	0.802431517789534\\
14	1.17827243566951\\
16	1.63755826680813\\
18	2.129290662867\\
20	2.57457397361345\\
22	2.93335101673211\\
24	3.20382225624571\\
26	3.39359755033641\\
28	3.52205310675343\\
30	3.60772641984677\\
32	3.66322779322015\\
34	3.70256456771849\\
36	3.73069755761775\\
38	3.75199877266366\\
40	3.76846932838218\\
};

\addplot [color=mycolor3, mark=diamond, mark options={solid, mycolor3}]
  table[row sep=crcr , y expr=\thisrowno{1}*\fac]{%
6	0.134105826726215\\
8	0.318428833101854\\
10	0.536804283551119\\
12	0.841355766219459\\
14	1.24668923655244\\
16	1.74793341054214\\
18	2.28827929305564\\
20	2.76733815456074\\
22	3.15733216968143\\
24	3.44186776323094\\
26	3.63474144279907\\
28	3.75806396563936\\
30	3.83679851556898\\
32	3.88495413069748\\
34	3.91326900392768\\
36	3.93232607446992\\
38	3.94623440496076\\
40	3.95582459749751\\
};


\addplot [color=mycolor2]
  table[row sep=crcr , y expr=\thisrowno{1}*\fac]{%
6	1.01469084069011\\
8	1.4210681632986\\
10	1.86071000079526\\
12	2.29541872272838\\
14	2.66298422987976\\
16	2.92864733084898\\
18	3.07075849147382\\
20	3.13518231837004\\
22	3.15204602866642\\
24	3.15412875456801\\
26	3.15420040186797\\
28	3.15420055390041\\
30	3.15420055390041\\
32	3.15420055390041\\
34	3.15420055390041\\
36	3.15420055390041\\
38	3.15420055390041\\
40	3.15420055390041\\
};

\addplot [color=mycolor2, mark=square, mark options={solid, mycolor2}]
  table[row sep=crcr , y expr=\thisrowno{1}*\fac]{%
6	0.974801569859215\\
8	1.36563144579934\\
10	1.78709803839326\\
12	2.2059769113785\\
14	2.57134870222725\\
16	2.83722419733626\\
18	3.0015668124996\\
20	3.08301628521574\\
22	3.12256638191279\\
24	3.13886341839329\\
26	3.14549404682931\\
28	3.14856755298797\\
30	3.15007321825299\\
32	3.15079978145955\\
34	3.15114442867528\\
36	3.15131916604351\\
38	3.15140390262712\\
40	3.15139971748658\\
};

\addplot [color=mycolor2, mark=O, mark options={solid, mycolor2}]
  table[row sep=crcr , y expr=\thisrowno{1}*\fac , y expr=\thisrowno{1}*\fac]{%
6	0.992398116928802\\
8	1.39101262082775\\
10	1.82081974576027\\
12	2.24787916269818\\
14	2.61377942808338\\
16	2.88646084157695\\
18	3.03667070291154\\
20	3.11358130942573\\
22	3.14370137060556\\
24	3.15170674528052\\
26	3.15359883991295\\
28	3.15402111041645\\
30	3.15415111368603\\
32	3.15417111083076\\
34	3.15419132297552\\
36	3.15419386619784\\
38	3.15419631037979\\
40	3.15419723371143\\
};

\addplot [color=mycolor2, mark=triangle, mark options={solid, mycolor2}]
  table[row sep=crcr , y expr=\thisrowno{1}*\fac]{%
6	0.974820900514243\\
8	1.36570457086056\\
10	1.78722766547578\\
12	2.20600929943679\\
14	2.57166544348797\\
16	2.83799698075448\\
18	3.00269718217651\\
20	3.0849080712994\\
22	3.12480912609414\\
24	3.1409598105072\\
26	3.14749964730845\\
28	3.15035065179121\\
30	3.15163446202757\\
32	3.15233572367285\\
34	3.15265873398537\\
36	3.15287172392781\\
38	3.15303666832911\\
40	3.15314359065086\\
};

\addplot [color=mycolor2, mark=diamond, mark options={solid, mycolor2}]
  table[row sep=crcr , y expr=\thisrowno{1}*\fac]{%
6	0.992385233224439\\
8	1.39099011159633\\
10	1.82087825769518\\
12	2.24790119604452\\
14	2.61394810351106\\
16	2.88684302925525\\
18	3.03723889251336\\
20	3.11436025806283\\
22	3.14437842057605\\
24	3.15215867830731\\
26	3.15380953393366\\
28	3.15413521108165\\
30	3.15418534550661\\
32	3.15419387118562\\
34	3.15420030118328\\
36	3.15420055390041\\
38	3.15420055390041\\
40	3.15420055390041\\
};


\addplot [color=mycolor1]
  table[row sep=crcr , y expr=\thisrowno{1}*\fac]{%
6	1.47038977951384\\
8	1.65374140354845\\
10	1.73010202138814\\
12	1.7452288054424\\
14	1.74617712536418\\
16	1.74619037242917\\
18	1.74619037242917\\
20	1.74619037242917\\
22	1.74619037242917\\
24	1.74619037242917\\
26	1.74619037242917\\
28	1.74619037242917\\
30	1.74619037242917\\
32	1.74619037242917\\
34	1.74619037242917\\
36	1.74619037242917\\
38	1.74619037242917\\
40	1.74619037242917\\
};

\addplot [color=mycolor1, mark=square, mark options={solid, mycolor1}]
  table[row sep=crcr , y expr=\thisrowno{1}*\fac]{%
6	1.44689804518872\\
8	1.6366827159626\\
10	1.72286686971644\\
12	1.74397732064866\\
14	1.74610716714469\\
16	1.74619037242917\\
18	1.74619037242917\\
20	1.74619037242917\\
22	1.74619037242917\\
24	1.74619037242917\\
26	1.74619037242917\\
28	1.74619037242917\\
30	1.74619037242917\\
32	1.74619037242917\\
34	1.74619037242917\\
36	1.74619037242917\\
38	1.74619037242917\\
40	1.74619037242917\\
};

\addplot [color=mycolor1, mark=o, mark options={solid, mycolor1}]
  table[row sep=crcr , y expr=\thisrowno{1}*\fac]{%
6	1.45851983350368\\
8	1.64527281587346\\
10	1.72670236260651\\
12	1.74470195922897\\
14	1.74615934451929\\
16	1.74619037242917\\
18	1.74619037242917\\
20	1.74619037242917\\
22	1.74619037242917\\
24	1.74619037242917\\
26	1.74619037242917\\
28	1.74619037242917\\
30	1.74619037242917\\
32	1.74619037242917\\
34	1.74619037242917\\
36	1.74619037242917\\
38	1.74619037242917\\
40	1.74619037242917\\
};

\addplot [color=mycolor1, mark=triangle, mark options={solid, mycolor1}]
  table[row sep=crcr , y expr=\thisrowno{1}*\fac]{%
6	1.44690444226245\\
8	1.63669101995861\\
10	1.72288106475856\\
12	1.74397601769344\\
14	1.74610869079713\\
16	1.74619037242917\\
18	1.74619037242917\\
20	1.74619037242917\\
22	1.74619037242917\\
24	1.74619037242917\\
26	1.74619037242917\\
28	1.74619037242917\\
30	1.74619037242917\\
32	1.74619037242917\\
34	1.74619037242917\\
36	1.74619037242917\\
38	1.74619037242917\\
40	1.74619037242917\\
};

\addplot [color=mycolor1, mark=diamond, mark options={solid, mycolor1}]
  table[row sep=crcr , y expr=\thisrowno{1}*\fac]{%
6	1.45852849862352\\
8	1.64527344743228\\
10	1.72670876561266\\
12	1.74470302284155\\
14	1.74615765156273\\
16	1.74619037242917\\
18	1.74619037242917\\
20	1.74619037242917\\
22	1.74619037242917\\
24	1.74619037242917\\
26	1.74619037242917\\
28	1.74619037242917\\
30	1.74619037242917\\
32	1.74619037242917\\
34	1.74619037242917\\
36	1.74619037242917\\
38	1.74619037242917\\
40	1.74619037242917\\
};

\end{axis}


\end{tikzpicture}%